# Self-similarity properties in a queuing network model

D.V. Lande, A.A. Snarskii,
*National Technical University of Ukraine "Kyiv Polytechnic Institute"*

In this paper a model of subscriber telephone network based on the concept of cellular automata is elaborated [1]. Some fractal properties inherent in the model are revealed that vary depending on parameters assigning its operation rules. The main advantage of the model in question is its compatibility with algorithmic methods - a finite set of formal rules, assigned on a finite set of elements (cells), allows precise realization in the form of algorithms.

It should be noted that simulation of queuing networks, in particular, simulation of communication networks by means of cellular automata is a common approach [2, 3], at the same time, revealing self-similarity effects in the simplest model considered below allows to take a fresh view of the fractal nature of teletraffic.

A system of cellular automata is a combination of similar cells definitely interconnected. The state of each cell is determined by the state of neighbouring cells, Thus, the state of the $j$-th cell at time moment $t+1$ is determined as follows:

$$y_j(t+1) = F\left(y_j(t), O(j), t\right),$$

where $F$ is a certain rule that can be expressed, for example, in terms of the Boolean algebra, and $O(j)$ is a set of neighbouring cells (neighbourhood).

Each cell of the system of cellular automata that corresponds to a model of closed queuing network will be considered as a telephone subscriber. It is supposed that each subscriber at random time moments can give a telephone call only to his near acquaintances that correspond to eight cells from the subscriber's neighbourhood, i.e. Moore's neighbourhood is used [4] – the cell $y_{i,j}$ will have its cell neighbours: $y_{i-1,j-1}$, $y_{i-1,j}$, $y_{i-1,j+1}$, $y_{i,j-1}$, $y_{i,j+1}$, $y_{i+1,j-1}$, $y_{i+1,j}$, $y_{i+1,j+1}$. The time distribution of calls to the nearest neighbours for each subscriber is exponential with parameter $\lambda$. If at the moment of call the addressee's telephone is free, connection takes place and conversation occurs, the time of which also obeys the exponential distribution (with parameter $\mu$). If at the moment of call the addressee's telephone is busy, the denial of service takes place. During the next cycle of cellular automata system the subscriber again makes an attempt to connect a random neighbour (not necessarily the same). If the subscriber's telephone proves to be free (in standby state), while he was called, the standby time is over, and the telephone goes over to a busy state for the same period as the telephone of calling party.

For visualization of the model (Fig. 1) the cells that are in "busy" state will become red (a darker colour), and free cells – blue (a lighter colour). Inside the cells there are figures shown – the number of cycles to service completion (positive number), and the number of cycles to next call (negative number).

In the model presented there are three parameters that can vary – the size of cellular automata field, $\lambda$ and $\mu$. All examples are illustrated by a relatively small

field – 15x15 cells, however, as investigations showed, the increase in field size has almost no effect on the model dynamics.

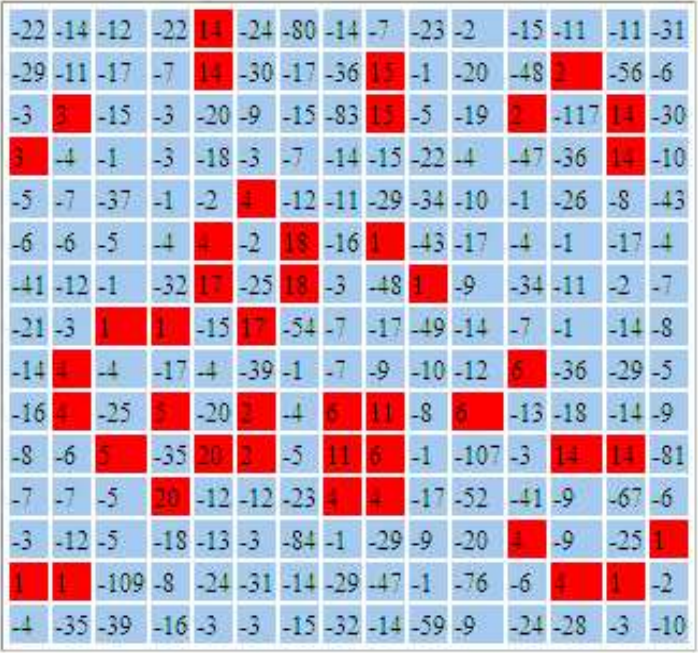

*Fig. 1. Cellular automata system simulating telephone network*

The output parameter of model is the number of cells that are in service, i.e. in "busy" state at current moment. The dynamics of changes in the output parameters for the values $\lambda=0.07$ and $\mu=0.03$ is shown in Fig. 2. It is obvious that at given output parameters the number of cells in service is constantly varied about some average value depending on these parameters. In so doing, the state of stabilization is not achieved, i.e. dispersion of the resulting series of observations does not tend to zero, moreover, it turned to be a constant value.

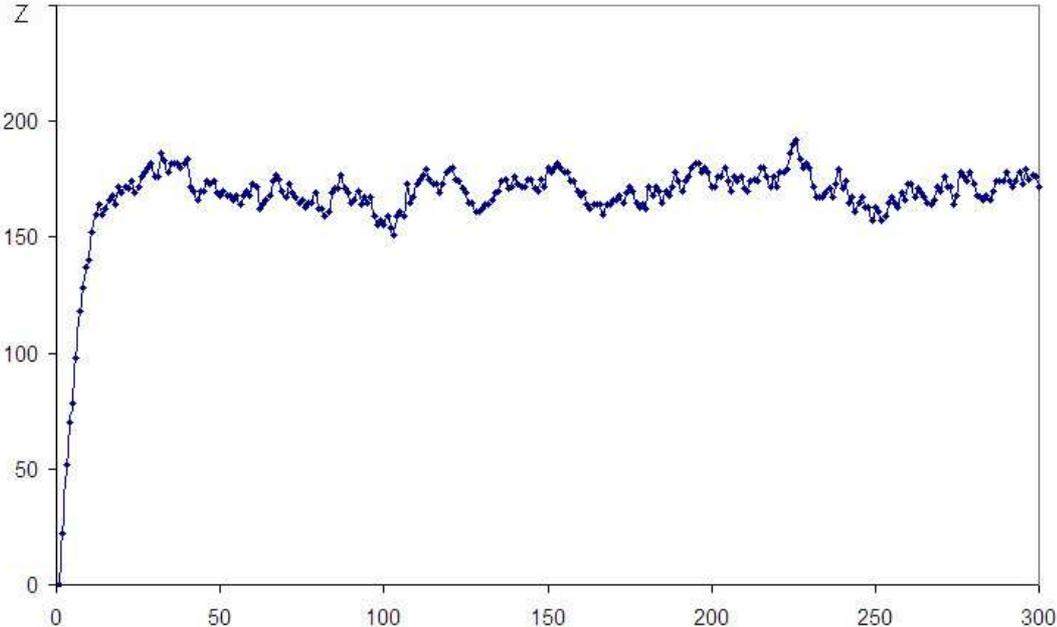

*Fig. 2. The number of cells (ordinate axis) in "busy" state Z versus the number of cycle (abscissa axis) at $\lambda=0.07$, $\mu=0.03$*

In contrast to dispersion, the average values of a series of observations Z (the number of boxes in "busy" state for different model cycles) for different values of parameters $\lambda$ and $\mu$ have a sufficiently regular behaviour (Fig. 3). The plot is interpolated by a simple relation:

$$Z = C \frac{\lambda}{1+\lambda} \frac{1+\mu}{\mu},$$

where $Z$ is a number of cells in service state, and $C$ is scaling constant depending on field size.

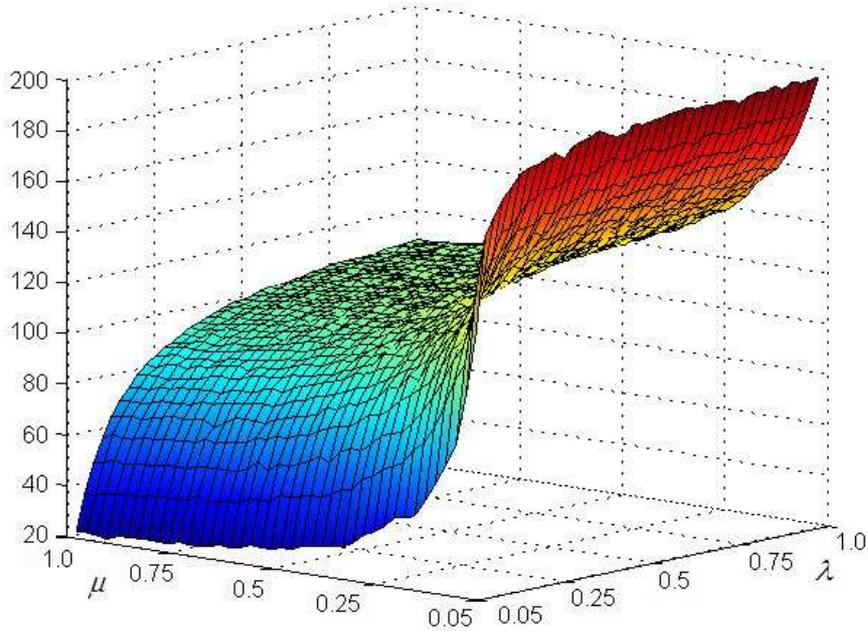

*Fig. 3. The average number of cells in "busy" state versus the values of model parameters*

The behaviour of Z series presented in Fig.2 gives grounds to suppose that we deal with a determined chaos, that this series can possess fractal properties, at least, self-similarity. This idea was verified by means of the so-called $R/S$-analysis [5]. In this case $R$ is a "spread" of the respective series of observations (assuming that one cycle of our model corresponds, for example, to 1 second, the series can be called temporal), and $S$ is a standard deviation. Let us explain how the spread value R is calculated. For the temporal series $F(n)$, $n = 1, ..., N$, we calculated the average value:

$$\langle F \rangle_N = \frac{1}{N} \sum_{n=1}^{N} F(n),$$

a series of accumulated values:

$$X(n,N) = \sum_{i=1}^{n} \left( F(i) - \langle F \rangle_N \right),$$

following which, direct measurement of spread is made:

$$R(N) = \max_{i \leq n \leq N} X(n,N) - \min_{i \leq n \leq N} X(n,N).$$

J.Hurst experimentally discovered the following to be valid for temporal series possessing the property of self-similarity: $R/S = (N/2)^H$, where $H$ is called the Hurst factor. The presence of this power dependence substantiates fractal properties of the temporal series. Fig. 4 shows a plot of $R/S$ versus cycle number in our investigated model for certain values of parameters $\lambda$ and $\mu$. The fact that this plot is well approximated by a straight line on a log-log scale gives grounds to suppose the presence of fractal properties in a model under consideration.

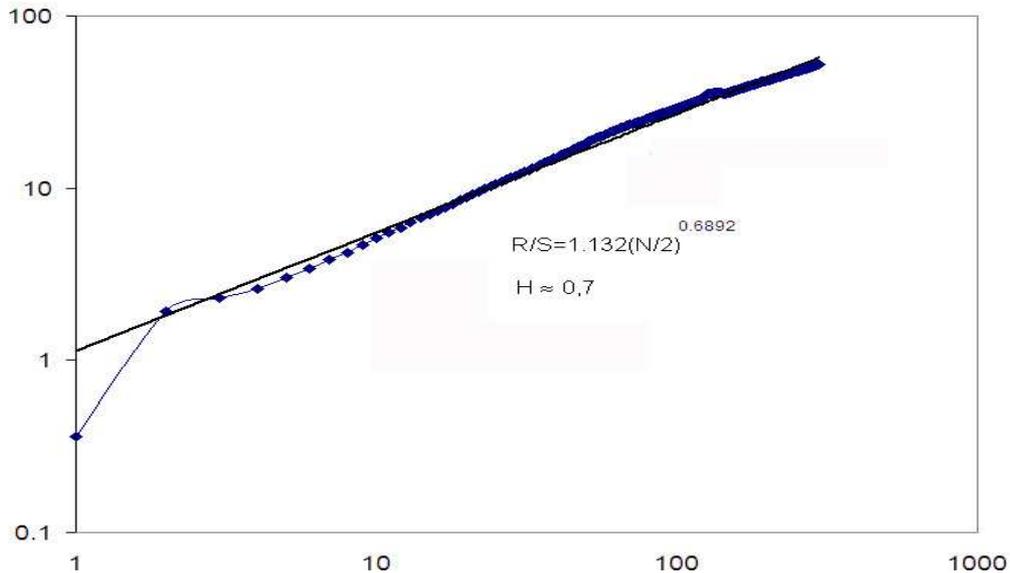

*Fig. 4. $R/S$ (ordinate axis) versus cycle number (abscissa axis) for parameter values $\lambda$ =0.07, $\mu$ =0.03 on a log scale*

The Hurst factor is known to be a measure of persistence, i.e. process disposition to trends. The value of $H > ½$ means that the process dynamics having certain direction in the past will most probably involve continued motion in the same direction. If $H < ½$, then it is predicted that the process will change its direction. $H = ½$ means uncertainty, for example, such value takes place for the Brownian motion. In the referred example the value of the Hurst factor is $H \approx 0.69$ (Fig. 4), which testifies to process persistence.

However, not for all series of observations the Hurst factor proved to be so "convincing"; at some $\lambda$ and $\mu$ values one could observe quite a chaotic behaviour of a series with the value of $H \approx ½$. Calculations were performed for different $\lambda$ and $\mu$ values, in order to clear out the character of self-similarity of $Z$ series. The data averaged already by 40 realizations (Fig. 5) testify to a stable regularity of change in the Hurst factor (the self-similarity level, respectively) as a function of $\lambda$ and $\mu$.

The investigations performed explain the self-similarity of teletraffic in some data communication networks [6] (for certain $\lambda$ and $\mu$ values ) on a simple and reproducible model, in particular, when the average standby time and connection duration is increased (in real systems it is first of all related to transfer of multimedia information), the Hurst factor grows, explaining the increased effect of teletraffic self-similarity in real networks [7].

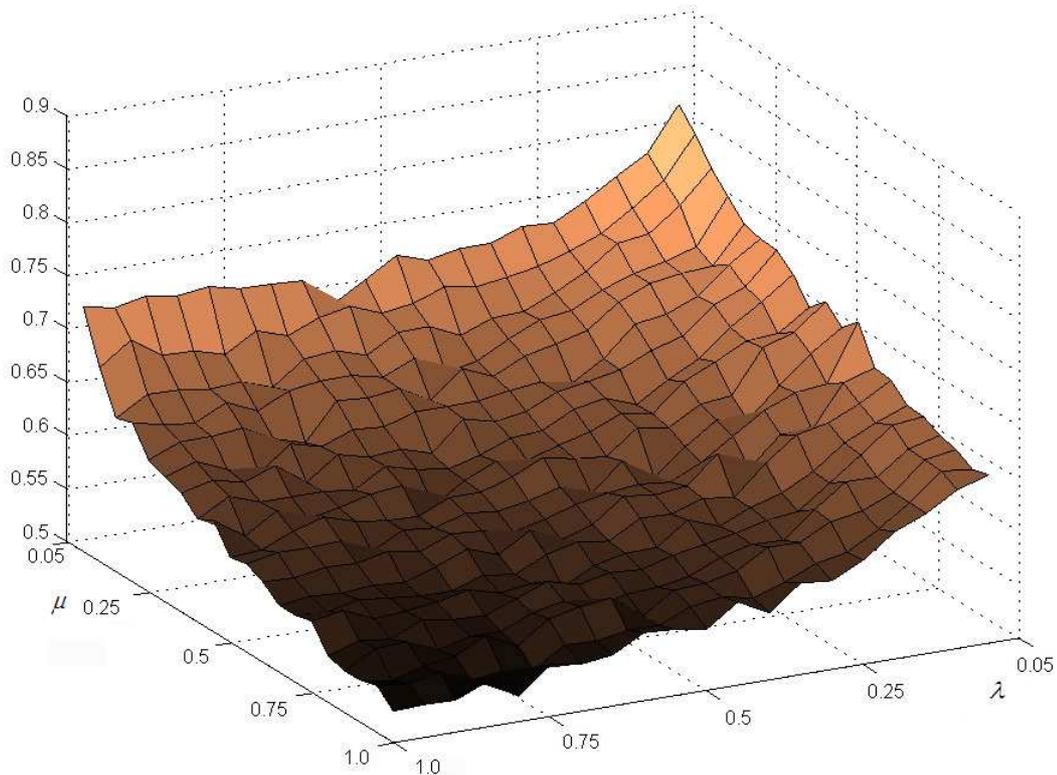

*Fig. 5. The plot of the Hurst factor versus the values of model parameters averaged by 40 realizations*